\documentstyle[epsfig,mnras_cite]{mn}
\title{3D weak lensing}
\author[Alan Heavens]{Alan Heavens\thanks{afh@roe.ac.uk} \\
Institute for Astronomy, University of Edinburgh, Royal
Observatory, Blackford Hill, Edinburgh, EH9 3HJ,UK}

\newcommand{\be}{\begin{equation}}  \newcommand{\ee}{\end{equation}}
  \newcommand{\ba}{\begin{eqnarray}}
\newcommand{\ea}{\end{eqnarray}}

\newcommand{\bb}{{\bf b}} \newcommand{\bg}{{\bf g}}  
  
  \newcommand{\br}{{\bf r}}
\newcommand{\bs}{{\bf s}}
  
\newcommand{\bx}{{\bf x}}  

\newcommand{\hn}{{\hat{\bf n}}}

\newcommand{\stp}{\sqrt{{2\over \pi}}}
\newcommand{\slf}{{1\over 2}\sqrt{{(\ell+2)!\over (\ell-2)!}}}

\def\gs{\mathrel{\raise1.16pt\hbox{$>$}\kern-7.0pt %
\lower3.06pt\hbox{{$\scriptstyle \sim$}}}}         %
\def\ls{\mathrel{\raise1.16pt\hbox{$<$}\kern-7.0pt %
\lower3.06pt\hbox{{$\scriptstyle \sim$}}}}         %

\begin{document}

\maketitle

\begin{abstract}
I propose an analysis method, based on spin-spherical harmonics
and spherical Bessel functions, for large-scale weak lensing
surveys which have source distance information through photometric
redshifts.  I show that the distance information can
significantly reduce statistical errors on cosmological
parameters; in particular, 3D lensing analysis offers excellent
prospects for constraining the equation of state of the vacuum
energy which dominates the energy density of the Universe.  I
show that the ratio of pressure to energy density could be
determined to an accuracy of $\sim 1\%$ or better. Having distance
information also offers significant advantages in the control of
systematic effects such as the intrinsic alignment of galaxies.
The case for obtaining photometric redshifts is therefore
compelling. A signal-to-noise eigenmode analysis of the modes
shows that the modes with highest signal-to-noise correspond
quite closely to ignoring the redshift information, but there is
significant extra information from a few radial modes.  These
modes are generally long-wavelength, suggesting that useful
information can be gleaned even if the photometric redshifts are
relatively inaccurate.
\end{abstract}


\title{3D weak lensing}

\section{Introduction}

As a direct probe of the mass distribution, gravitational lensing
is an excellent tool for cosmological parameter estimation,
complementing microwave background studies.  Many of the
cosmological parameters have been determined well from the
microwave background, most recently and most accurately by the
WMAP satellite \cite{Spergel03}.  There are some parameters which
are difficult to constrain from the microwave background alone,
the most significant example being the equation of state parameter
($w=p/\rho c^2$) of the vacuum energy.  An accurate determination
of $w$ and its time-dependence is a major goal in the post-WMAP
era, as it can put constraints on the nature of the vacuum energy.
The direct mass dependence also offers advantages over studies of
large-scale structure, which have to make assumptions about, or
deduce, the relationship between galaxy and mass distributions on
large scales \cite{Verde02}. One of the most useful
manifestations of gravitational lensing by intervening matter is
the alignment of nearby images on the sky. Detection of dark
matter on large scales through such `cosmic shear' measurements
has recently been shown to be feasible, and cosmological
constraints from the first detections have now been made (e.g.
\pcite{BRE00,vW00,Brownetal02}).  To date, the cosmic shear
analyses have concentrated on using essentially only
two-dimensional shear information;  in surveys where
(photometric) redshifts of the sources are known, they are used
to determine the redshift distribution only, and are not used
individually. Recently, though, 3D information has begun to be
employed, mostly through what I would describe as $2{1\over 2}$D
analysis:  the sources are divided into slices at different
distances, based on their photometric redshifts, and a 2D
analysis is done on each slice. This can be valuable in detecting
the presence of clusters of galaxies
(\pcite{Wittman01,Wittman02,Tayloretal03}). At a statistical
level, \scite{Hu99} has shown that there is some extra
information on cosmological parameters which can be gained by
dividing the sample this way.  The extra information depends on
the parameter under consideration:  for the amplitude of the
matter power spectrum, there are significant gains when the source
population is split into two, but little is gained by further,
fine subdivisions.  For $w$, the gains are much larger, as the
constraints from 2D lensing are weak. Finally, \scite{Taylor02}
has shown how the lensing equations can be inverted to deduce the
gravitational potential directly, and this offers exciting
possibilities of reconstruction of the 3D mass density field from
shear and convergence data \cite{BaconTaylor02}.
\scite{HuKeeton02} have shown how this inversion can be done by
matrix methods for discrete shear data in cells.

The purpose of this paper is to demonstrate how the full 3D
information about the shear can be used.  In particular, I
investigate how source photometric redshift information can be
used in a statistical analysis to determine the mass power
spectrum and the equation of state of the vacuum energy. In
particular, we are interested in how much extra information on
cosmological parameters can be gleaned from a 3D shear map, to
see whether the reduction in statistical errors justifies the
extra observing time required to obtain photometric redshifts for
the sources.

Reducing statistical errors is, of course, not the only reason for
wishing to have photometric redshifts; weak shear studies are
beset by a number of systematic errors, and these may eventually
dominate the error budget as surveys improve in scope and
quality.  The most obvious of these is the redshift distribution
of the sources - ignorance of the precise distribution is already
a source of significant error in current surveys (see e.g.
\pcite{BTHD02}).  At a different level, the analysis of cosmic
shear usually makes the assumption that the source galaxy
ellipticities are uncorrelated, and this may not be true in
detail, due to tidal effects during the galaxy formation process
(\pcite{HRH00,CM00,CNPT01,CKB01,Mackey02,Jing02}). The level of
this effect is not well known, but it is thought to be large
enough to be an important source of systematic error in
low-redshift lensing surveys (\pcite{BTHD02,HeyHea03}). By its
nature, tidal effects are important for galaxies which are close
together in three dimensions, so photometric redshifts can be
used to remove or downweight nearby galaxy pairs in shear
correlation estimates, for example \scite{HeyHea03},
\scite{KingSchneider02}.   These studies show that the systematic
effects of intrinsic alignments can be essentially completely
removed, at the expense of an increase in shot noise. Other
physical effects which could affect weak lensing studies can also
be removed this way, for example the effects of source clustering
if one uses number counts as a measure of magnification
\cite{BTP95}. Furthermore, with many (typically 5) images of
galaxies available from a photometric redshift lensing survey,
there is scope for better shape measurement, independent lensing
studies and so on.

We see that there is a powerful case for obtaining photometric
redshifts for lensing surveys on the basis of removing systematic
errors.  Having obtained distance information for the sources, it
seems sensible to use it in the statistical analysis, as it must
reduce the error bars on the parameters which one wants to
estimate.  In this paper, I propose a method, based on spherical
Bessel functions and spin-spherical harmonics, for analysing such
surveys.

I find that the 3D information can reduce the error on the
amplitude of the matter power spectrum by a factor $\sim 2$, which
broadly speaking compensates for the extra time taken to obtain
photometric redshifts.  The error on the vacuum equation of state
is encouraging: with 3D information $w$ could be determined to an
accuracy of $<1\%$.   These results assume that other
cosmological parameters are known accurately, most plausibly from
microwave background studies.

The paper is arranged as follows.  Section 2 describes the
method, section 3 discusses the expected errors, section 4
performs a signal-to-noise analysis of the modes, which turns out
to be very informative, and section 5 presents the conclusions.

\section{Method}

\subsection{Spherical coordinates and spherical Bessel functions}

Spherical coordinates are natural for the description of fields on the
full sky, but there are reasons for using them even if the sky coverage is
relatively small.  The survey specification is normally a combination
of sky coverage (specified in terms of angular coordinates $\theta,
\varphi$), and depth (related to $r$).  In addition, photometric
errors introduce errors in $r$, and the lensing potential is related
to the Newtonian potential by a radial integral.  Having said this,
the spherical harmonic expansion becomes cumbersome at high $\ell$,
and at some point a Fourier description on the flat sky becomes an
attractive approximation to make.  We choose to expand radially in
spherical Bessel functions, as these (when combined with spherical
harmonics) are eigenfunctions of the Laplace operator, which leads to
a very simple relationship between the coefficients of the
gravitational potential and those of the overdensity field, since
these are related by Poisson's equation.

\subsection{Transformation of scalar and shear fields}

The spherical harmonic transform of a scalar field $f(\br)$ is
defined by
\begin{equation}
f_{\ell m}(k) \equiv \sqrt{2\over \pi} \,
\int\,d^3\br f(\br)\,j_\ell(kr)Y_\ell^{m*}(\Omega)
\end{equation}
where $j_\ell(z)$ is a spherical Bessel function, $Y$ a spherical
harmonic, $k$ is a wavenumber, $\ell$ is a positive integer and
$m=-\ell, \ldots \ell$. The shear field, comprising $\gamma_1$
and $\gamma_2$, can be written as components of a tensor, and
expanded in terms of tensor spherical harmonics, as in
\scite{Taylor02}, or spin-spherical harmonics $_sY_\ell^m$ as in
\scite{OH02}.  We will use the latter notation, generalised to 3D:
\begin{eqnarray}
\gamma_1(\br) \pm \ i\gamma_2(\br) &=& \stp \sum_{\ell m}\slf\\\nonumber
& & \int dk k^2 \, \gamma_{\ell m}(k) _{\pm 2} Y_\ell^m(\hn)j_\ell(k\,r).
\end{eqnarray}
In principle the coefficients $\gamma_{\ell m}(k)$ could depend on
whether the $+$ or $-$ sign is taken, but they are the same, and are
related to the transform of the lensing potential $\phi(\br)$ by
\begin{equation}
\gamma_{\ell m}(k) = {1\over 2}\sqrt{(\ell+2)!\over (\ell-2)!}\,
\phi_{\ell m}(k)
\end{equation}
\cite{Taylor02}.  The lensing potential is related to the
gravitational potential $\Phi(\br)$ by a radial integral (e.g. \pcite{BS})
\begin{equation}
\phi(\br) =
{2\over c^2} \int_0^r dr' \left[{f_K(r)-f_K(r')\over f_K(r)f_K(r')}\right]
\Phi(\br').
\end{equation}
$f_K(r)d\psi$ is the dimensionless transverse comoving separation
for points separated by an angle $d\psi$.  The Robertson-Walker
metric may be written $ds^2 = c^2 dt^2 - R^2(t)\left[dr^2 +
f_K^2(r)d\psi^2\right]$, and $f_K(r)$ takes the values $\sin r,\
r,\ \sinh r$ for curvature values $k=1,0,-1$.

$\Phi(\br)$ is related to the overdensity field $\delta(\br) \equiv
\delta\rho(\br)/\bar\rho$ by Poisson's equation
\begin{equation}
\nabla^2\Phi = {3\Omega_m H_0^2\over 2 a(t)}\delta,
\label{PoissonReal}
\end{equation}
where $\Omega_m$ is the matter density parameter, $H_0$ is the
Hubble constant and $a(t)=1/(1+z)$ is the scale factor. $\delta$
itself is not a homogeneous field, because it evolves with time,
and hence with distance from the observer through the light
travel time.  In the linear regime, we can describe the growth by
a universal growth rate $g(t)$ (see, for example \scite{AD02} for
quintessence models), so that $\delta = \delta^0 g(t)$, where
$\delta^0(\br)$ is the overdensity field, extrapolated using
linear theory to the present day, and which is homogeneous. Thus
$\Phi(\br) = \Phi^{0}g(r)/a(r)$. Note that for an Einstein-de
Sitter universe, $g(r)/a(r)=1$, and that for the `concordance'
model of matter and vacuum density parameters $\Omega_m=0.3$,
$\Omega_v=0.7$, $g(r)/a(r)$ declines from 1.28 at high redshift
to unity today.

The transform of the lensing potential is therefore related to that of the
present-day potential field by
\begin{equation}
\phi_{\ell m}(k) = {1\over
c^2}\int_0^\infty \, dk' k'^2 \eta_\ell(k,k')\Phi^0_{\ell m}(k')
\label{Potential}
\end{equation}
where
\begin{eqnarray}
\eta_\ell(k,k') &\equiv& {4\over \pi}\int_0^\infty dr f_K(r)
j_\ell(k r) \nonumber\\
& & \int_0^r dr' j_\ell(k'r')\left[{f_K(r)-f_K(r')\over
f_K(r')}\right] {g(r')\over a(r')}.
\label{Eta}
\end{eqnarray}
These equations allow us to relate the coefficients of the shear
field, $\gamma_{\ell m}(k)$ to the underlying linear, present-day
overdensity field, $\delta_{\ell m}(k)$, since Poisson's equation
implies
\begin{equation}
\Phi^0_{\ell m}(k) = -{3\Omega_m H_0^2\over 2 k^2}\delta^0_{\ell
m}(k). \label{Poisson}
\end{equation}

\subsection{Discrete estimator}

The data which we have to hand are estimates of the shear field at 3D
positions in space, so it makes some sense to transform the data directly.
The radial coordinate is generally not known
accurately, but is typically given by a photometric redshift which has
an error which may be $\sigma_z \sim 0.02-0.1$ or more.  We denote this radial
coordinate by $s$, and the true coordinate by $r$.  They are related
by a conditional probability, which we will take to be a gaussian
\begin{equation}
p(s|r)ds = {1\over \sqrt{2\pi}\sigma_z}\exp\left[-{(z_s-z_r)^2\over
2 \sigma_z^2}\right] dz_s
\end{equation}
where $z_{r,s}$ are the redshifts associated with distance coordinates
$r$ and $s$.  Note that $\sigma_z$ can vary with $z$.

The estimate of the shear is obtained from the complex ellipticity
\cite{BS}
\begin{equation}
e = e^s+2\gamma
\end{equation}
where $e^s$ is the intrinsic source ellipticity, whose mean is assumed
to be zero, so $\hat \gamma = e/2$ is an unbiased, albeit noisy,
estimate of the shear field.  The shear estimates have covariance
properties
\begin{equation}
\langle \hat\gamma_a \hat\gamma_b^*\rangle = \langle\gamma_a
\gamma_b^*\rangle + {1\over 4}\langle e_a^s
e_b^{s*}\rangle
\end{equation}
for galaxies $a$ and $b$.  We denote the variance of $e$ by
$\sigma_e^2$, typically $\ls 0.1$ \cite{Hudson98}.  The second
term isnon-zero if galaxies are intrinsically aligned (see e.g.
\pcite{HRH00}). I ignore this term, as with photometric
information, its effect can be removed by modifying the analysis
to remove close pairs in three dimensions
(\pcite{HeyHea03,KingSchneider02}).  Thus we have noisy estimates
$\hat\gamma$ of the shear field at imprecise positions with
distances $s$ given by photometric redshifts.

An obvious set of quantities to use for a harmonic description of the
data is simply to do a discrete transform of the measurements which
are to hand.  We consider the following quantities:
\begin{equation}
_\pm \hat g_{\ell m}(k) \equiv \stp \sum_{\rm galaxies\ g}\left(\hat
\gamma_{1g} \pm i \hat\gamma_{2g} \right) j_\ell(k s_g) \ _{\pm
2}Y_{\ell}^m(\hn_g).
\label{estimate}
\end{equation}
$\ _\pm \hat g_{\ell m}(k)$ are estimates of the quantities
\begin{equation}
_\pm g_{\ell m}(k) = \stp \int d^3\bs\,
n(\bs) \gamma_{\pm}(\br) j_\ell(k s) \ _{\pm
2}Y_{\ell}^m(\hn),
\end{equation}
where $n(\br)$ is the number density, and $\gamma_{\pm}\equiv
\gamma_1\pm i\gamma_2$.  It is straightforward to include a
non-uniform weighting of the galaxies if desired.   This may
improve the errors on parameter estimation, but is not explored
in this paper.

Note that in this equation the shear field is evaluated at the
true position $\br$, whereas the estimate (\ref{estimate})
involves the distance $s$ estimated from the photometric
redshift.  These are related by the conditional probability
$p(s|r)$, leading to an average value of the expansion
coefficients
\begin{eqnarray}
_\pm \bar g_{\ell m}(k) &=& \stp \int d^3\bs \int dr\, p(s|r)\nonumber\\
& &n(\bs) \gamma_{\pm}(\br) j_\ell(k s) \ _{\pm 2}Y_{\ell}^m(\hn).
\label{g1}\end{eqnarray} With the photometric redshift smoothing,
$n(\bs)$ is heavily smoothed, so we can approximate it by the
smoothed number density at $\br$, $n_s(\br)$.  For deep surveys,
the angular clustering is small, so we can ignore clustering, and
approximate this by the average number density, which normally
separates into a radial part and an angular selection:
\begin{equation}
n_s(\br) = \bar n(r) M(\hn)
\end{equation}
Normally $M=0$ (unobserved sky) or $M=1$ (in survey), although
more complicated forms are possible, if some parts of the survey
are partially sampled.

If we denote the spin-spherical harmonic transform of $n_s(\br)
\gamma_\pm(\br)$ by $_\pm h_{\ell m}(k)$, i.e.
\begin{equation}
n_s(\br) \gamma_\pm(\br) = \stp \sum_{\ell m} \int dk\, k^2 \, _\pm h_{\ell
m}(k) j_\ell(k r) _{\pm 2}Y_\ell^m(\hn)
\end{equation}
we find that (\ref{g1}) may be written
\begin{eqnarray}
_\pm \bar g_{\ell m}(k) &=& {2\over \pi} \int ds f_K^2(s) d \hn \,dr\, p(s|r)
\nonumber\\ &&
\sum_{\ell' m'} \int dk' k'^2 \ _\pm h_{\ell' m'}(k)  j_\ell'(k' r)
j_\ell(k s) \nonumber\\
& & \ _{\pm 2}Y_{\ell'}^{m'}(\hn)\ _{\pm 2}Y_\ell^{m*}(\hn)M^2(\hn).
\label{gbar}
\end{eqnarray}
It is convenient to define angular mixing matrices by
\begin{equation}
_\pm W_{\ell\ell'}^{mm'} \equiv \int d\hn  \ _{\pm
2}Y_{\ell'}^{m'}(\hn)
\ _{\pm 2}Y_\ell^{m*}(\hn)M^2(\hn).
\label{Wll}
\end{equation}
For all-sky coverage,$\ _\pm
W_{\ell\ell'}^{mm'}=\delta^K_{\ell\ell'}\delta^K_{mm'}$, where
$\delta^K$ is the Kronecker delta symbol.  Using $W$, we can simplify
(\ref{gbar}) to
\begin{equation}
_\pm \bar g_{\ell m}(k) = \sum_{\ell' m'} \int dk' k'^2\ _\pm
W_{\ell\ell'}^{mm'} Z_{\ell\ell'}(k,k') _\pm h_{\ell' m'}(k')
\end{equation}
where
\begin{equation}
 Z_{\ell\ell'}(k,k') \equiv  {2\over \pi} \int ds f_K^2(s) dr\, p(s|r)
j_{\ell'}(k' r) j_\ell(k s). \label{Zll}
\end{equation}
$_\pm h_{\ell m}(k)$ may be calculated by direct substitution of the
expansion of $\gamma_{\pm}$, yielding
\begin{equation}
_\pm h_{\ell m}(k) = \int dk' k'^2 M_{\ell\ell}(k,k')\ _\pm \gamma_{\ell m}(k')
\end{equation}
where
\begin{equation}
M_{\ell\ell'}(k,k')\equiv {2\over \pi} \int dr f_K^2(r) j_{\ell'}(k' r)
j_\ell(k r) \bar n(r).
\label{Mll}
\end{equation}
(This equation is more general than necessary at this stage, with
two indices $\ell$ and $\ell'$,  but $M_{\ell\ell'}$ will appear
in its general form in the shot noise below). It is convenient to
define a continuous form of the Einstein summation convention
indicating integration over wavenumber:
\begin{equation}
A(k,k')B(k',k'') \equiv \int dk' k'^2 A(k,k') B(k',k'').
\end{equation}
With this notation, we can write the shear expansion coefficients in
terms of the present-day, linear density field coefficients as follows
(from (\ref{Wll}), (\ref{Zll}), (\ref{Eta}), (\ref{Poisson}) and
(\ref{Potential})):
\begin{eqnarray}
_\pm \bar g_{\ell m}(k) &=& -A\sum_{\ell'm'} \sqrt{{(\ell'+2)!\over
(\ell'-2)!}}
\ _\pm W_{\ell\ell'}^{mm'}
Z_{\ell\ell'}(k,k') \nonumber\\
&&M_{\ell\ell}(k',\tilde k)
{\eta_{\ell'}(\tilde
k,\tilde k')\over \tilde k'^2} \delta^{0}_{\ell' m'}(\tilde k')
\label{gdrelation}
\end{eqnarray}
where $A\equiv 3\Omega_m H_0^2/(4c^2)$.  Note that the sums over
$\ell'$ and $m'$ are made explicitly.

For all-sky surveys, the average values of $_\pm g_{\ell m}(k)$
are zero; information about cosmological parameters comes from
their variance (or, more precisely, their covariance).

\subsection{Covariance matrix of $_\pm \bar g_{\ell m}(k)$}

Since the present-day linear power spectrum is a homogeneous field,
its covariance matrix is diagonal:
\begin{equation}
\langle  \delta^{0}_{\ell m}(\tilde k) \delta^{0*}_{\ell' m'}(\tilde
k')\rangle = {P_\delta^0(k)\over
k^2}\delta^D(k-k')\delta^K_{\ell\ell'}\delta^K_{mm'}
\end{equation}
where $\delta^D$ is the Dirac delta function, and $P_\delta^0(k)$ is
the present-day linear overdensity power spectrum.

The signal part of the covariance matrix is then
\begin{eqnarray}
\langle _\pm \bar g_{\ell m}(k) _\pm \bar g_{\ell' m'}^*(k')
\rangle &=& A^2\sum_{\tilde\ell\tilde m} {(\tilde\ell+2)!\over
(\tilde\ell-2)!} \ _\pm W_{\ell\tilde\ell}^{m\tilde m} \ _\pm
W_{\ell'\tilde\ell}^{m'\tilde m*}\nonumber\\
& & \!\!\!\!\!\!\!\!\!\!\!\!
Z_{\ell\tilde\ell}(k,k_1)Z_{\ell'\tilde\ell}(k',k_2)
M_{\tilde\ell\tilde\ell}(k_1,k_3)
M_{\tilde\ell\tilde\ell}(k_2,k_4) \nonumber \\ &
&\!\!\!\!\!\!\!\!\!\!\!\!
{\eta_{\tilde\ell}(k_3,k_5)\eta_{\tilde\ell}(k_4,k_5) \over k_5^4}
P_\delta^{0}(k_5). \label{Signal}
\end{eqnarray}
Note that if coverage is all-sky, the $\,_\pm W$ matrices become
delta functions and the covariance matrix simplifies considerably
to
\begin{eqnarray}
\langle _\pm \bar g_{\ell m}(k) _\pm \bar g_{\ell' m'}^*(k')
\rangle &=& A^2 {(\ell+2)!\over (\ell-2)!}
Z_{\ell\ell}(k,k_1)Z_{\ell\ell}(k',k_2)  \nonumber
\\ & &
\!\!\!\!\!\!\!\!\!\! \!\!\!\!\!\!\!\!\!\! \!\!\!\!\!\!\!\!\!\!
\!\!\!\!\!\!\!\!\!\! \!\!\!\!\!\!\!\!\!\! \!\!\!\!\!\!\!\!\!\!
\!\!\!\!\!\!\!\!\!\! \!\!\!\!\! M_{\ell\ell}(k_1,k_3)
M_{\ell\ell}(k_2,k_4) {\eta_{\ell}(k_3,k_5)\eta_{\ell}(k_4,k_5)
\over k_5^4} P_\delta^{0}(k_5)\delta^K_{\ell\ell'}\delta^K_{mm'}.
\label{SignalAllSky}
\end{eqnarray}

\subsection{Shot Noise}

The shot noise can be computed via standard methods (e.g.
\pcite{Peebles}), by dividing the volume into cells $i$
containing $n_i=0$ or $1$ galaxy.
\begin{equation}
_\pm \hat g_{\ell m}(k) \equiv \stp \sum_{{\rm cells\ }i} n_i\hat
\gamma_{\pm i} j_\ell(k s_i) \ _{\pm
2}Y_{\ell}^m(\hn_i).
\end{equation}
so
\begin{eqnarray}
\langle\ _\pm \hat g_{\ell m}(k)\ _\pm \hat g_{\ell'
m'}^*(k')\rangle &=& {2\over \pi} \sum_{ij} \langle n_i n_j \hat
\gamma_{\pm i}\ \hat\gamma_{\pm j}^* \rangle j_\ell(k s_i)
\nonumber\\
& &\!\!\!\!\!\!\!\!\!\!\!\! j_\ell(k s_j) \ _{\pm
2}Y_{\ell}^{m*}(\hn_i)\ _{\pm 2}Y_{\ell}^m(\hn_j).
\end{eqnarray}
Assuming the shear field is uncorrelated with the presence or absence
of a source galaxy, $ \langle n_i n_j \hat \gamma_{\pm i}\
\hat \gamma_{\pm j}^* \rangle = \langle n_i n_j\rangle \langle \hat
\gamma_{\pm i}\hat\gamma_{\pm j}^* \rangle$.  Ignoring the small
correlations in the smoothed density field,
\begin{eqnarray}
\langle n_i n_j\rangle &=& \langle n_i \rangle \qquad (i=j)\nonumber\\
& &   \langle n_i \rangle  \langle n_j \rangle \qquad (i\ne j).
\end{eqnarray}
The shot noise comes from the term $i=j$, for which
\begin{equation}
\langle |\hat \gamma_{\pm i}|^2\rangle = \langle
|\gamma_{\pm i}|^2 \rangle + {\sigma_e^2\over 4}
\end{equation}
where the second term will dominate for weak lensing.  Thus the shot
noise may be written
\begin{eqnarray}
\langle\ _\pm \hat g_{\ell m}(k)\ _\pm \hat g_{\ell'
m'}^*(k')\rangle_{SN} &=& {\sigma_e^2 \over 2 \pi} \int ds f_K^2(s) \bar
n(s)    \nonumber\\ & & j_\ell(k s)  j_{\ell'}(k' s)
\ _\pm W_{\ell\ell'}^{mm'}
\nonumber\\
&=& {\sigma_e^2\over 4} M_{\ell\ell'}(k,k')  \ _\pm W_{\ell\ell'}^{mm'},
\label{Noise}
\end{eqnarray}
using the definition (\ref{Mll}).

\section{Estimation of cosmological parameters}

Various parts of this analysis are dependent on cosmological
parameters: the matter power spectrum $P_\delta^0(k)$; the components
of the metric $r(z)$ and $f_K[r(z)]$; the growth rate of perturbations
$g(t)$.  These parameters ($\{\theta_\alpha\}$) may be estimated from
the data using likelihood methods.  Assuming uniform priors for the
parameters, the maximum a posteriori probability for the parameters is
given by the maximum likelihood solution.  For large-scale modes we
use a gaussian likelihood
\begin{equation}
2\ln L(\bg|\{\theta_\alpha\}) = {\rm constant}-\det(C)-\bg \cdot
C^{-1} \cdot \bg
\end{equation}
where $C=S+N$ is the covariance matrix, given by signal and noise
terms (\ref{Signal}) and (\ref{Noise}).  Note that the average values
of \bg\ (the set of $\ _\pm \hat g_{\ell m}(k)$) is zero, so the
information on the parameters comes from the dependence of the signal
part of $C$.  i.e. we adjust the parameters until the {\em covariance
} of the model matches that of the data.  This was the approach of
\scite{HT95} and \scite{BHT95} in analysis of large-scale galaxy data.

Considering the vector \bg\ as the data set has some advantages over
more traditional methods such as the shear correlation function or the
shear variance.  The main one is that it is linear in the shear, so
the covariance matrix is quadratic, and readily calculable.  Quadratic
estimators such as the shear correlation function have 4th order
covariances, which can be cumbersome to calculate, even in gaussian,
linear theory.  We do, however, make an assumption that the covariance
matrix is gaussian, and it requires numerical simulations to establish
for which ranges of $k$ and $\ell$ this is a good approximation.

\subsection{Expected errors on cosmological parameters - the Fisher
matrix}

For a given experimental setup, the Fisher matrix gives the best
errors to expect, provided that the likelihood surface near the
peak is adequately approximated by a multivariate gaussian.  We
illustrate the effectiveness of a 3D analysis by analysing an
all-sky survey with the following details, and compare some of
the results with a 2D analysis of the same data. For smaller
surveys, a reasonable approximation is to scale the errors by
$f_{sky}^{-1/2}$, where $f_{sky}$ is the fraction of sky
observed.   We assume that the average number density is $n(z)
\propto z^2 \exp\left[-(z/z_*)^{1.5}\right]$ (cf \cite{BE93}),
where $z_*=z_m/1.412$ and $z_m$ is the median redshift of the
survey, which we take to be 1.  We assume a source density of 30
or 100 per square arcminute, errors on the ellipticity of 0.2 or
0.3, and a $\Lambda CDM$ model with matter and vacuum density
parameters of $\Omega_m=0.3$, $\Omega_v=0.7$.

The Fisher matrix is the expectation value of the second derivative of
the $\ln L$ with respect to the parameters $\theta_\alpha$:
\begin{equation}
F_{\alpha\beta} = -\left\langle {\partial^2 \ln L\over \partial
\theta_\alpha \partial\theta_\beta}\right\rangle
\end{equation}
and the marginal error on parameter $\theta_\alpha$ is
$\sqrt{(F^{-1})_{\alpha\alpha}}$ \cite{TTH97}.  If the means of the data
are fixed, the Fisher matrix can be calculated from the covariance
matrix and its derivatives by \cite{TTH97}
\begin{equation}
F_{\alpha\beta} = {1\over 2}{\rm
Trace}\left[C^{-1}C_{,\alpha}C^{-1}C_{,\beta}\right].
\label{Fisher1}
\end{equation}
For an all-sky survey, this simplifies, because modes with
different $\ell$ and $m$ are uncorrelated ($\,_\pm
W_{\ell\ell'}^{mm'}=\delta^K_{\ell\ell'}\delta^K_{mm'}$), so,
defining $C^\ell$ as the covariance matrix for the modes with
harmonic $\ell$ (and different $k$),
\begin{equation}
F_{\alpha\beta} = {1\over 2}\sum_\ell (2\ell+1){\rm
Trace}\left[(C^\ell)^{-1}C^\ell_{,\alpha}(C^\ell)^{-1}C^\ell_{,\beta}\right].
\label{Fisher2}
\end{equation}
For illustration, we have considered modes up to $\ell=100$, and
100 $k$ modes spaced equally between $k=0.001$ and $0.1\,
h$Mpc$^{-1}$. These should be safely in the linear regime.  The
$\ell$ range can be extended, but at some point the linear
approximation will break down, and the likelihood expression will
deviate from a gaussian.  The former may be tackled by using a
nonlinear power spectrum, such as proposed by \scite{PD96}; the
latter will need to be assessed by computer simulation.

\subsection{Power spectrum amplitude}

\begin{figure}
\centerline{\psfig{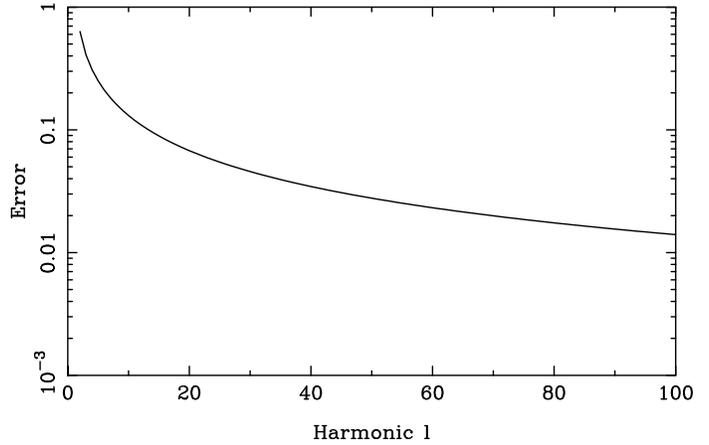}}
\caption{Fractional error on the amplitude of the linear matter
density power spectrum, assuming all other parameters are known,
from a 2D analysis of the survey, where individual source distance
information is ignored.  The figure shows the improvement as more
harmonics are added up to $\ell=100$.  100 radial wavenumber
modes are considered between  $k=0.001$ and $0.1\, h$Mpc$^{-1}$.
Illustrated model is $\Lambda CDM$, with 100 source galaxies per
square arcminute, and an ellipticity error of 0.2.  Plot is
cumulative, showing the reduction in error as more modes are
included.\label{Sigma2D}}
\end{figure}

\begin{figure}
\centerline{\psfig{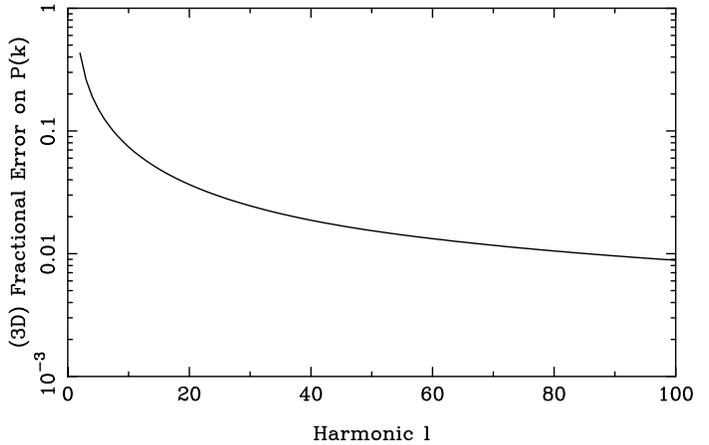}}
\caption{As Fig. \ref{Sigma2D}, except that a 3D analysis is used, with
photometric redshifts having an r.m.s. error of $\sigma_z=0.02$. \label{Sigma3D}}
\end{figure}

Fig. \ref{Sigma2D} and
\ref{Sigma3D} show the improvement in the error on the fractional
amplitude of the power spectrum as the number of $\ell$ modes is
increased up to $\ell=100$.  For a 2D analysis, ignoring the distance
information altogether, except for assuming $n(z)$ is known, the error
is 1.4\% for $\ell\le 100$.  Using the photometric redshift distance
information improves this to $0.9\%$, assuming, perhaps
optimistically, that the photometric redshift error is $0.02$.

The improvement depends on the characteristics of the survey.
Generally speaking, the better the lensing survey (higher number
density of source galaxies, smaller ellipticity errors), the
better the 3D analysis does in comparison with 2D, at least on
the scales shown here.  The reason is simply that the 3D modes
are generally noisy, whereas the large-scale 2D modes are not.
Improving the survey brings in more effective 3D modes which have
good signal-to-noise.  The effective number of good radial modes
is investigated further in the Karhunen-Loeve analysis of section
4.
\begin{figure}
\centerline{\psfig{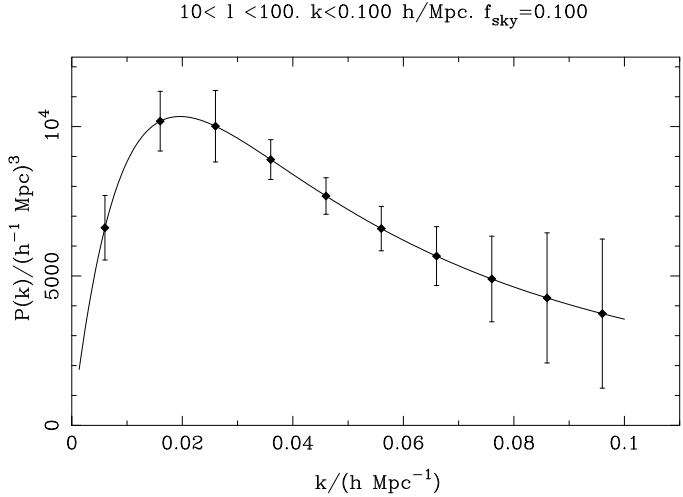}}
\caption{Band power estimates for a survey of 10\% of the sky.
The source number density assumed is 30 per square arcminute. The
error bars are marginalised over the other bands. Note that the
increase in the error at high $k$ is a result of our truncating
the analysis at $\ell=100$. Increasing this would provide more
information on the power spectrum at high $k$. \label{Band1}}
\end{figure}

\begin{figure}
\centerline{\psfig{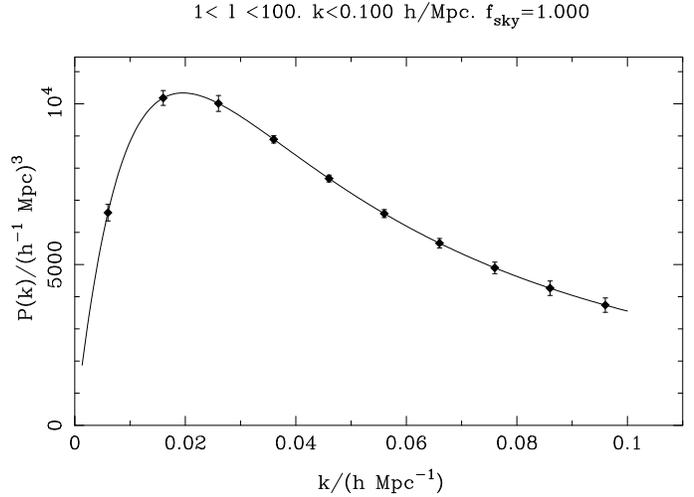}}
\caption{As Fig. \ref{Band1}, except survey is all-sky, intrinsic
ellipticity dispersion is 0.2.\label{Band2}}
\end{figure}

The Fisher analysis can be extended to calculate the errors arising if we
estimate band powers from the 3D lensing data.  The Fisher matrix is calculated as before,
and its inverse used to estimate the errors on the band powers.  Fig. \ref{Band1} shows what may
be achieved realistically for a survey of 10\% of the sky, with a median redshift of 1.  Fig.
\ref{Band2} is for an optimistic all-sky survey and optimistic errors.

\subsection{Vacuum Energy equation of state}

The most exciting prospect of 3D lensing analysis is to measure
the equation of state parameter, $w$, of the vacuum energy,
defined in terms of its pressure and energy density by $p_v = w
\rho_v c^2$.   The equation of state of the vacuum energy
influences the lensing signal in two ways. Firstly, the growth
rate of perturbations differs; secondly, the distance-redshift
relation is changed.  I use the growth rate from \scite{AD02} for
quintessence models, and the distance-redshift relation is
obtained from the integral
\begin{equation}
r = \int_0^z\,{dz'\over H(z')}
\end{equation}
where for flat models
$H(z)=H_0[(1-\Omega_v)(1+z)^3+\Omega_v(1+z)^{3(1+w)}]^{-1/2}$ is
the Hubble parameter in terms of the present day Hubble parameter
$H_0$, and $\Omega_v$ is the present vacuum energy density
parameter.

The sensitivity to $w$ can then be calculated by computing the
Fisher matrix (scalar), equation (\ref{Fisher2}), where the
derivative of the covariance matrix is obtained by finite
differencing.  Note that this figure is optimistic in the sense
that the errors it produces assume that all other parameters are
known, not altogether unrealistic with the success of microwave
background experiments.  On the other hand, $\ell=40$ is a
reasonably large scale ($\sim 5^\circ$), and there is more
information available at higher $\ell$, so 1\% accuracy is
probably achievable.

\begin{figure}
\centerline{\psfig{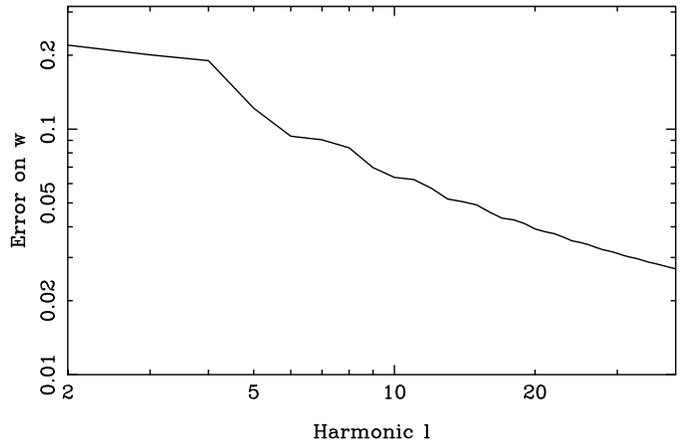}}
\caption{The expected error on $w$ for a survey covering 10\% of
the sky to a median redshift of 1, with 100 galaxies per square
arcminute, and a photometric redshift error of 0.02, and an
ellipticity error of 0.3.  The plot is cumulative to harmonic
$\ell$.  In practice for partial sky coverage the low-$\ell$
modes will not be accessible, but this will hardly change the
results.}
\end{figure}

\section{Karhunen-Lo\` eve analysis}

For all-sky surveys, the coupling of modes is only between those
of the same $\ell$ and $m$, so the data analysis task is
relatively straightforward, as each set of $\ell$ and $m$ can be
analysed separately.  For partial sky coverage, the modes get
mixed, and the covariance matrix gets very large.  Data
compression techniques such as signal-to-noise eigenmodes,
Karhunen-Lo\` eve (KL) transformations and the like can be
extremely valuable in reducing the size of the data set whilst
having minimal impact on the error bars on the recovered
parameters.  These have been used extensively in studies of
large-scale structure and the microwave background (refs).

\begin{figure}
\centerline{\psfig{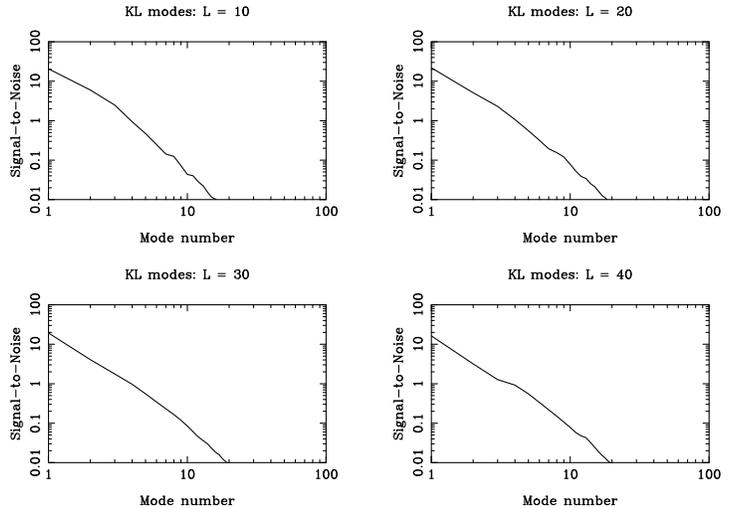}}
\caption{Signal-to-noise eigenmode analysis of the sample survey,
for a few of the low-order angular modes.  We see that there are
about 4 useful linear combinations of the radial $k$
modes.\label{KL}}
\end{figure}

The procedure is detailed, for example, in \scite{TTH97}, and
involves finding linear combinations of the data which are
uncorrelated and which are ordered in decreasing order of the
information which they yield on a parameter.  This involves
solving a generalised eigenvalue problem
\begin{equation}
C_{,\alpha}{\bb} = \lambda C \bb
\end{equation}
where $\theta_\alpha$ is the parameter of interest.  The
eigenvalues $\lambda$ quantify how much information each linear
combination $\bb\cdot \bx$ of the data $\bx$ provides, and very
often this declines sharply with mode number, so the dataset can
be compressed substantially without significantly increasing the
error bar. Here, we calculate the KL modes appropriate for the
(natural log of the) amplitude of the matter power spectrum, in
which case $C_{,\alpha}=S$. The problem then is equivalent to
finding signal-to-noise eigenmodes, e.g.\scite{VS96}. Fig.
\ref{KL} shows the signal-to-noise of the best modes for each of
the low-order multipoles.  We see that there are roughly 4 useful
modes with $S/N>1$.  In other words, there is some extra
information on the power spectrum coming from the 3D information,
but the added accuracy is limited - the analysis had 100 radial
wavenumbers. This is broadly in agreement with the findings of
\scite{Hu99}, who found that division of the source population
into two, on the basis of redshift, improved the error bars on
the power spectrum by a factor of 2, but that further division
yielded little extra accuracy.

\begin{figure}
\centerline{\psfig{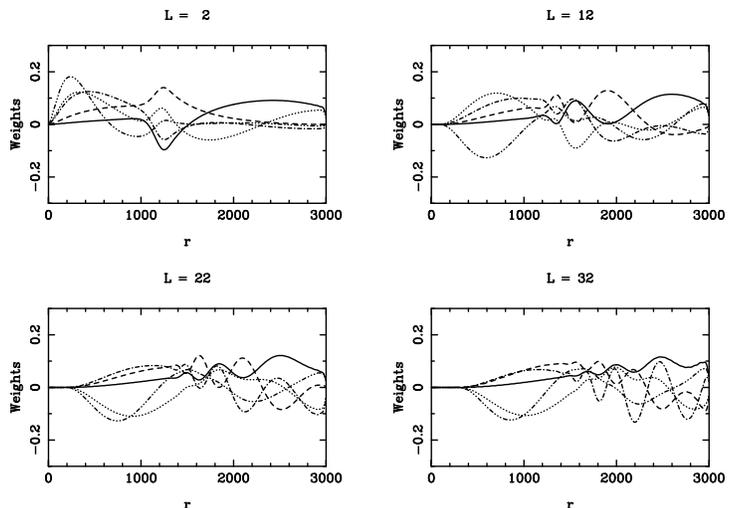}}
\caption{KL modes corresponding to the low-order multipoles.  In
each case the best mode is solid, and the dashed, dot-dashed,
dotted and the dash-dot-dot-dotted lines show the next 4 modes in
decreasing order of usefulness.  Note that a) the best couple of modes
largely ignore the distance information, as in a
2D survey, and b) the best modes are long-wavelength - accurate
photometric redshifts may not be required.\label{KLmodes}}
\end{figure}

It is quite instructive to look at the KL modes themselves.
Fig. \ref{KLmodes} shows the best radial KL modes corresponding with
the low-order angular multipoles.  We see several interesting
features.  Firstly, the best two modes between them weight the data
reasonably uniformly, apart from a
drop-off at small distances.  This latter behaviour is expected, as
the lensing is ineffective at low redshift.  The near-uniform
weighting suggests that a 2D survey is not too bad - it corresponds to
equal weighting, and matches quite well the single best mode of the KL
analysis.   For other parameters, this may not hold.

The other interesting and encouraging feature of these graphs is that
the good modes are long-wavelength, suggesting that having precise
photometric redshifts is not necessary to get most of the information
present in the data.

\section{Conclusions}

In this paper, I have shown how photometric redshifts may be used
to perform a full 3D statistical analysis of the shear field in
weak lensing surveys.  With photometric redshifts, one has an
estimate of the lensing shear field at a the (estimated)
positions of the source galaxies in three dimensions, and there
is no particular reason to throw this away by ignoring the
distance information or by dividing the source galaxy population
into shells.

Determining the properties of the vacuum energy, which makes up
$\sim 70\%$ of the energy density of the Universe, is one the most
important current goals in cosmology.   These properties are not
easy to determine accurately using the microwave background
alone, which has been so successful in recent years in pinning
down other cosmological parameters (e.g. \pcite{Spergel03}). I
show in this paper that 3D lensing analysis offers the
possibility of high-precision measurement of the equation of
state parameter $w$ (defined such that the vacuum has
$p_v=w\rho_v c^2)$, with an accuracy of 1\% being a realistic
possibility.  With higher angular resolution than investigated
here, the prospects of getting good constraints on $w(z)$ are
good, and the possibilities of testing specific predictions of
$w(z)$ from models of the vacuum energy are excellent.  For other
parameters, such as the amplitude of the matter power spectrum,
3D information reduces the error bars by modest factors.  A
signal-to-noise eigenmode analysis suggests that there are a few
radial modes which are useful for this purpose. This analysis
also shows that the high signal-to-noise modes have little
high-frequency structure, so even modestly accurate photometric
redshifts can be useful.

In addition to the advantages in reducing the statistical error
on parameters, photometric redshifts also allow elimination of a
number of possible systematic effects, arising from physical
processes, which could otherwise eventually limit the accuracy of
weak lensing studies.  For cosmic shear measurements in
particular, the dominant physical systematic may be the intrinsic
alignment of galaxies arising from tidal forces during and after
formation (\pcite{HRH00,CM00,CNPT01,CKB01,Mackey02,Jing02}).
\scite{HeyHea03} and \scite{KingSchneider02} showed how
photometric redshifts can be used to remove nearby pairs from the
analysis and essentially remove the intrinsic alignment
contamination to high accuracy.  The advantages in reduced
statistical and systematic errors make a compelling case for
obtaining photometric redshifts for cosmic shear surveys.

\section{Acknowledgments}

I would like to thank Andy Taylor, Meghan Gray, David Bacon and
Catherine Heymans for useful discussions.



\end{document}